\documentclass[openacc]{rstransa}

\begin{document}

\title{New futures for cosmological models}

\author{
L. Fern\'andez-Jambrina$^{1}$ and R. Lazkoz$^{2}$}

\address{$^{1}$Matem\'atica e Inform\'atica Aplicadas, Avenida de la 
Memoria 4, 28040-Madrid, Spain\\
$^{2}$F\'\i sica Te\'orica, Facultad de Ciencia y Tecnolog\'\i a,
Universidad del Pa\'\i s Vasco, Apdo.  644, E-48080 Bilbao, Spain
}
\subject{Theoretical physics}


\corres{Ruth Lazkoz\\
\email{ruth.lazkoz@ehu.es}}

\begin{abstract}
The discovery of accelerated expansion of the universe opened the
possibility of new scenarios for the doom of our spacetime, besides
aeternal expansion and a final contraction.  In this paper we review
the chances which may await our universe.  In particular, there are
new possible singular fates (sudden singularities, big rip...), but
there also other evolutions which cannot be considered as singular.
In addition to this, some of the singular fates are not strong enough
in the sense that the spacetime can be extended beyond the
singularity.  For deriving our results we make use of generalised
power and asymptotic expansions of the scale factor of the universe.
\end{abstract}


\begin{fmtext}
\end{fmtext}
\maketitle

\section{Introduction}

The discovery of accelerated expansion of our universe has posed many
new issues in gravitational theory.  One of them is that the final
fate is not constrained to aeternal expansion or recollapse into a Big
Crunch singularity.  One by one, quite a few names have been added to
such list of fates (Big Rip, Little Rip, sudden singularities\ldots).
Some of them are singular (strong or weak), some are not.  In this
paper we would like to review this issue and provide a thorough
classification of them.

The issue of defining what a singularity is in General Relativity is
definitely not a simple one \cite{geroch,HE,beem}.  Most field
theories are defined in a fixed background spacetime manifold, for
instance Minkowski spacetime. A field is then singular when
it diverges at a certain event in the background spacetime. This is
what happens, for instance, with the electric Coulombian field due to
a single charged particle, which is singular at the location of the
particle.  Many of such singularities are removed on considering
quantum field theories.

In General Relativity the problem is a bit more involved and plural.
We cannot consider a field such as the metric of the spacetime as
singular if it diverges at some set of coordinates, since the theory
is invariant under changes of coordinates.  There are no privileged
coordinates and it could happen that the metric does not diverge in
other system of coordinates.  This is what happens, for instance, for
the horizon of the Schwarzschild metric, which appears to be singular
in some coordinates, but not in Eddington-Finkelstein or Kruskal
coordinates.  Furthermore, if the gravitational field is singular, the
spacetime manifold would be singular too.  We cannot say that the
gravitational field is singular at one \emph{event} as in other field
theories, since a singularity of the gravitational field is a 
singularity of the spacetime itself and hence we cannot talk about 
events or points of the manifold.

One could relate singularities to divergences of the curvature
scalars.  These are named scalar polynomial curvature singularities
(s.p. curvature singularities).  This is a common test for the
appearance of singularities, since it is easy to check whether the
curvature scalars diverge, pointing out that we are leaving the 
spacetime manifold.

However, such polynomials do not fully characterise the curvature
tensor.  One can resort then to scalar polynomials in the derivatives
of the curvature tensor.  On the other hand, curvature tensor
components may diverge along a curve, though the scalar polynomials
remain finite.  A parallelly propagated (p.p) curvature singularity
appears when at least a component of the curvature tensor in a
parallelly propagated basis along a curve grows unboundedly.  These
p.p. curvature singularities include s.p. curvature singularities.

Another way of dealing with this is considering causal geodesics in
the manifold, that is, timelike and lightlike geodesics.  Free-falling
observers, subject just to gravitational interaction, follow causal
geodesics in General Relativity.  And these geodesics can be
parametrised using the proper time or affine parameter, which is
related to time as experienced by the free-falling observer in the
case of timelike geodesics.  

This provides us another way of defining singularities.  In principle
proper time should range from minus infinity to plus infinity.  If
this cannot be done, after a finite proper time our observer would be
out of the manifold and this behavior can be considered as a token of
a singularity.  A singularity appears in a manifold if a causal
geodesic is incomplete in this sense, but we must take care of having
a maximal extension for our spacetime (the horizon of Schwarschild
spacetime is not singular in Kruskal's extension, but it appears to be
singular in other charts).

This is one of the most common definitions of singularities and most theorems 
are referred to geodesic incompleteness. Of course this definition 
comprises curvature singularities, but there are spacetimes with 
incomplete geodesics which are not related to curvature singularities 
(geodesic imprisonment). This is what happens, for instance, in 
Taub-NUT spacetime.

One reason for considering geodesic completeness is that in Riemannian
geometry metric completeness (every Cauchy sequence is convergent) is
equivalent to geodesic completeness, though it is not so in Lorentzian
geometry.

Finally, one can extend the previous definition to non-geodesic curves
(accelerated observers) by introducing a generalised affine parameter
for them.  A spacetime is bundle complete (b-complete), and in this
sense singularity-free, if every finite-length curve has an endpoint
in the manifold. That is, an observer does not leave the spacetime 
manifold in a finite time.

In the next Section we derive and solve the geodesic equations for 
flat FLRW cosmological models. In Section~3 we review the concept of 
strength of a singularity and apply it to our models. Section~4 is 
devoted to power and asymptotic expansions of the deceleration 
parameter in order to find all possible singular 
behaviors, by relating the expansions to the energy-momentum content 
of the models. A thorough classification of the models in terms of 
their final behavior is provided in Section~5.

\section{Geodesics in cosmological models}

For dealing with cosmological models we consider 
Friedman-Lema\^{i}tre-Robinson-Walker (FLRW) homogeneous and
isotropic spacetimes.  We introduce coordinates $t,r,\theta,\phi$ with
the usual meaning and ranges, such that the metric can be written as
\begin{equation}ds^2=-dt^2+a^2(t)\left(dr^2+
r^2\left(d\theta^2+\sin^2\theta\,
d\phi^2\right)\right),\label{metric}\end{equation}in terms of the 
scale factor of the universe $a(t)$. We just consider spatially flat 
models for their match with observations, but the analysis can be 
done for non-flat models either.

Geodesics can be parametrised with their proper, afine or internal time 
$\tau$, such that $d\tau^{2}=-ds^{2}$. The velocity of the 
parametrisation of a geodesic would be then $u=(\dot t, \dot r, \dot 
\theta, \dot \phi)$, where the dot stands for derivation with respect 
to proper time. 

We may define a conserved quantity $\delta$,
\begin{equation}\delta:=-u\cdot u=\dot t^2-a^{2}(t)\left(\dot r^2+
r^2\left(\dot\theta^2+\sin^2\theta\, \dot\phi^2\right)\right),
\end{equation}
which is zero for lightlike geodesics, one for timelike
geodesics and minus one for spacelike geodesics.

Since the spacetime is homogeneous and isotropic, geodesics are 
straight lines and without losing generality we may take 
$\dot\theta=0$, $\dot\phi=0$. Furthermore, there is a conserved 
momentum associated to $\partial_{r}$ along such straight lines,
\[\pm P=u\cdot \partial_{r}=a^{2}(t)\dot r,\]
and this allows us to write down the geodesic equations in a simple 
fashion without resorting to calculation with Levi-Civita symbols,
\begin{subequations}\begin{eqnarray}\label{geods}
    \dot t&=&\sqrt{\delta +\frac{P^2}{a^2(t)}},\label{geods1}\\\dot
    r&=&\pm\frac 
    {P}{a^2(t)},\label{geods2}\end{eqnarray}\end{subequations}
where the sign of $\dot t$ is chosen as positive to deal with future-pointing 
geodesics.

As we see, there are three kinds of causal geodesics, namely 
lightlike ($\delta = 0$, $P\neq 0$), radial timelike ($\delta = 1$, $P\neq 0$) and
timelike comoving ones, ($\delta = 1$, $P=0$), which are the worldlines of a 
fluid of velocity $u$.

In order to go on with the analysis of geodesic behavior we need more 
information about the scale factor $a(t)$. Without much loss of 
generality, we may assume that it can be expanded around an event at  
$t_{0}$ \cite{visser,puiseux,modigravi},
\begin{equation}\label{puiseux}
    a(t)=c_{0}|t-t_{0}|^{\xi_{0}}+c_{1}|t-t_{1}|^{\xi_{1}}+\cdots\;,
\end{equation}
with real (not necessarily integer) and ordered exponents 
$\xi_{0}<\xi_{1}<\cdots$.

At first view we notice that the value of $\xi_{0}$ already conveys 
information about the possible singular behavior of the geodesic,
\begin{itemize}
\item If $\xi_{0}>0$, $a(t)$ vanishes at $t_{0}$ and we find a Big
Bang or Big Crunch singularity.

\item If $\xi_{0}=0$, $a(t)$ is finite at $t_{0}$ and we may 
encounter a singularity depending on whether $a(t)$ is analytical or 
there is a divergent derivative of the scale factor.

\item If $\xi_{0}<0$, $a(t)$ diverges at $t_{0}$ and we associate 
this with a Big Rip singularity.
\end{itemize}

There is another possibility which cannot be overlooked, corresponding
to an infinite value of the time coordinate $t$
\cite{hidden,grandrip,initial}.  This needs to be considered if there
are observers which may reach $t=\infty$ in finite proper time.  In
the case of lightlike geodesics we have
\[ \dot t=\frac{P}{a(t)},\qquad
\int_{t_{0}}^t a(t')\,dt'=P(\tau-\tau_{0}),\]
and hence $t=\infty$ is reached in finite proper time if the integral
\begin{equation}\label{conda}\int_{t}^\infty a(t')\,dt'\end{equation}
is finite for large $t$. 

For instance, if we allow an asymptotic behavior for the scale factor 
of the form \[a(t)\simeq c|t|^{\xi},\] for large $t$, we see that, 
according to (\ref{geods1}) $t$ diverges at finite proper
time $\tau_{0}$ if $\xi<-1$, the affine time of lightlike geodesics 
ranges from $\tau=-\infty$ ($t=0$) to
$\tau=\tau_{0}$ ($t=\infty$) and are future-incomplete. Since the 
curvature scalar polinomials vanish at infinity time coordinate, we 
find here a simple example of p.p. curvature singularities. 

\section{Strength of the singularities}

The concept of strength of singularities arises in the context of
gravitational collapse, not in cosmological frameworks.  Big Bang and
Bing Crunch singularities are strong in the sense that, as
respectively the origin and end of the universe, no observer should 
ellude them. This is not so in gravitational collapse and, as we 
shall see, with other types of singularities.

The concept of strength of a singularity requires considering finite
extended bodies instead of pointlike observers (causal geodesics).  A geodesic may be not
continued beyond a singularity, but one has to take into account if
tidal forces are capable of disrupting an extended body on approaching
the singularity \cite{ellis}.  If this is the case, the singularity is considered
strong.

There are several ways of conveying mathematical rigor to this
concept.  In these definitions the finite object is depicted as the
infinitesimal volume spanned by an orthonormal basis
parallelly-transported along the causal geodesic.  In Tipler's
definition \cite{tipler}, the singularity is strong if such volume
tends to zero at the singularity.  The main consequence is that the
spacetime may be extended beyond weak singularities and in this sense
a weak singularity cannot be considered the final fate.

However, according to the less restrictive Kr\'olak's definition
\cite{krolak}, the singularity is strong if the derivative of this
volume is just negative.

For checking the fulfillment of these definitions, there are simple
necessary and sufficient conditions \cite{clarke} which require
calculation of integrals of the curvature tensor along the incomplete
geodesic.

%

According to Tipler's definition, an incomplete lightlike geodesic
of velocity $u$ comes up a strong singularity at its affine time $\tau_{0}$ if and
only if the integral
\begin{equation}\label{suftipler}
 \int_{0}^{\tau}d\tau'\int_{0}^{\tau'}d\tau''R_{ij}u^{i}u^j
\end{equation}
diverges as $\tau$ approaches $\tau_{0}$. $R$ stands for the Ricci 
tensor.

On the other hand, according to Kr\'olak's definition, an incomplete lightlike geodesic
of velocity $u$ comes up a strong singularity at its affine time $\tau_{0}$ if and
only if the integral
\begin{equation}\label{sufkrolak}
 \int_{0}^{\tau}d\tau'R_{ij}u^{i}u^j
 \end{equation}
diverges as $\tau$ approaches $\tau_{0}$.

In our case, the integrand is readily written in terms of the scale 
factor and its derivatives,
\begin{equation}
R_{ij}u^iu^j\,d\tau=2P\left(\frac{a'^2}{a^3}-\frac{a''}{a^2}\right) dt
.\end{equation}

For timelike geodesics the conditions are the same, but they are just 
sufficient conditions, not necessary conditions.

For comoving geodesics, the integrand is just
\begin{equation}
R_{ij}u^i u^j\,d\tau=-\frac{3a''}{a}\,dt,\end{equation}
whereas for radial timelike geodesics is
\begin{equation}
R_{ij}u^i u^j\,d\tau=
\frac{-\frac{3a''}{a}+2P^2\left(\frac{a'^2}{a^4}-\frac{a''}{a^3}
\right)}{\sqrt{1+\frac{P^2}{a^2}}}dt.\end{equation}

These results applied to an expression for $a(t)$ in the form 
(\ref{puiseux}) readily allow to determine whether a singularity is 
weak or strong.

\section{Generalised power expansion of energy-momentum in 
cosmological models}

The energy-momentum content of FLRW spacetimes can be seen as a
perfect fluid of energy density $\rho(t)$ and pressure $p(t)$, which
thanks to the Einstein equations are readily written in terms of the
scale factor and its derivatives with respect to the time coordinate 
$t$ (Friedman equations),
\begin{equation}\label{flrw}\rho= \frac{3
a'^2}{a^2},\qquad p=-\frac{2 a''}{a}-\frac{
a'^2}{a^2}.\end{equation}

The energy density is much related to the Hubble ratio $H(t)$,
\begin{equation}H=\frac{a'}{a},\end{equation}
a measure of the expansion of the universe. 

The second derivative is usually expressed in terms of the 
deceleration parameter $q(t)$,
\begin{equation}q=-\frac{a a''}{ a'^2},\end{equation}
which is also closely related to the barotropic index of the
cosmological model $w(t)$, the quotient between pressure and
density,
\begin{equation}
w=\frac{p}{\rho}=-\frac{1}{3}-\frac{2}{3}\frac{a a''}{ a'^2}.\end{equation}

We shall see that it is useful to perform calculations using the 
deceleration parameter instead of the scale factor.

It is also common to define  a time coordinate $x=\ln a$,
\begin{equation}\frac{ x''}{ x'^2}=-(q+1),\end{equation}
which invites us to define the deviation $h(t)$ of the deceleration 
parameter from the  pure cosmological constant case \cite{grandrip},
\begin{equation}q(t)=-1+h(t).\end{equation}
	
Besides being physically meaningful, this definition allows us to 
reduce the order of Friedman equations,
\[h=-\frac{ x''}{ x'^2}=\left(\frac{1}{ 
x'}\right)^\cdot \Rightarrow  x'=\left(\int h\,dt 
+K_{1}\right)^{-1},\]
and to integrate the scale factor of the model,
\begin{equation}a(t)=\exp\left(\int\left(\int h(t)\,dt 
+K_{1}\right)^{-1}dt+K_{2}\right).\end{equation}

We may calculate explicitly the energy density and pressure of the 
model by fixing the integration limits,
\begin{equation}\rho(t)=3 x'(t)^2=3\left(\int_{t_{0}}^t h(t')\,dt'
+K_{1}\right)^{-2},\end{equation}
\begin{equation}p(t)=-2 x''(t)-3 x'(t)^2=\frac{2h(t)-3}{\left(
\displaystyle\int_{t_{0}}^t h(t')\,dt' 
+K_{1}\right)^{2}},\end{equation}
and interpreting the integration constants.

One constant can be absorbed as a global constant factor 
$a(t_{0})=\exp(K_{2})$, which is just the value of the scale factor 
nowadays,
\begin{equation}\label{scale}
a(t)=a(t_{0})\exp\left(\int_{t_{0}}^t\left(\int_{t_{0}}^{t''} h(t')\,dt' 
+K_{1}\right)^{-1}dt''\right).\end{equation}

The other constant may be interpreted in terms of the energy density,
$K_{1}=\sqrt{3}\rho(t_{0})^{-1/2}$, except in the case of infinite 
$\rho(t_{0})$, for which $K_{1}=0$, which is relevant for the 
appearance of singularities. For the sake of simplicity, we take 
$t_{0}=0$, $a(t_{0})=1$.

Without much loss of generality, we assume for our analysis of
singularities that the relevant function $h(t)$ can be expanded in 
real powers of $t$ around the event at $t=0$,
\begin{equation}\label{expandh}
h(t)=h_{0}t^{\eta_{0}}+ h_{1}t^{\eta_{1}}+ \cdots, \qquad
\eta_{0}<\eta_1<\cdots,\end{equation} where he have assumed positive
$t$, but the same analysis can be performed just exchanging $t$ for
$-t$.

The  scale factor, the energy density and the 
pressure behave at lowest order in $t$ as
\begin{equation}x(t)=\left\{\begin{array}{ll}\displaystyle-\frac{\eta_{0}+1}{\eta_{0} 
h_{0}}t^{-\eta_{0}}+\cdots &\textrm{if\ } -1\neq \eta_{0}\neq 0\\\\
\displaystyle \frac{1}{h_{0}}\int\frac{dt}{\ln|t|}+\cdots &\textrm{if\ } 
\eta_{0}=-1\\\\
\displaystyle\frac{\ln|t|}{h_{0}}+\cdots
&\textrm{if\ }\eta_{0}=0.
\end{array}\right.\end{equation}
\begin{equation}
\rho(t)=\left\{\begin{array}{ll}\displaystyle 3\left(\frac{\eta_{0}+1}{h_{0}}\right)^2t^{-2(\eta_{0}+1)}+\cdots
&\textrm{if\ }-1\neq \eta_{0}\neq 0\\\\\displaystyle 
\frac{3}{ h_{0}^2}\frac{1}{\ln^{2}|t|}+\cdots&\textrm{if\ }\eta_{0}=-1
\\\\\displaystyle
\frac{3t^{-2}}{h_{0}^2}
+\cdots 
&\textrm{if\ }\eta_{0}= 0,\end{array}\right.\end{equation}
\begin{equation}
	p(t)=\left\{\begin{array}{ll}\displaystyle
\frac{2(\eta_{0}+1)^2}{h_{0}}t^{-\eta_{0}-2}+\cdots 
&\textrm{if\ }-1\neq\eta_{0}<0\\\\\displaystyle 
\frac{2}{ h_{0}}\frac{1}{t\ln^{2}|t|}+\cdots &\textrm{if\ 
}\eta_{0}=-1\\ \\\displaystyle
\frac{2h_{0}-3}{h_{0}^2}t^{-2}
+\cdots  &
\textrm{if\ }\eta_{0}=0\\
\\ -\displaystyle
3\left(\frac{\eta_{0}+1}{h_{0}}\right)^2t^{-2(\eta_{0}+1)}+\cdots &
\textrm{if\ }\eta_{0}>0.
\end{array}\right.\end{equation}

Calculations are lengthy and one has to be careful in order to 
prevent overlooking subcases. From these expressions one can 
determine if the scale factor $a_{s}$, the energy density $\rho_{s}$, 
the pressure $p_{s}$ or even the barotropic index $w_{s}$ 
diverge or not at $t=0$, depending on the expansion (\ref{expandh}), 
which is what most classifications of singularities 
\cite{Nojiri:2005sx,IV,yurov,sesto} take into
consideration. The strength or weakness of the singularities can be 
determined with the results of the previous section.

We have summarised the possible cases 
in Table~\ref{tablw}, in terms of our power expansion. The last 
column enlarges classifications \cite{Nojiri:2005sx,IV,yurov,sesto} and shall be 
explained later.
\begin{table}[h]
   \begin{tabular}{cccccccc}
   \hline
   ${\eta_{0}}$ &$a_{s}$ & $\rho_{s}$ &$p_{s}$ & $w_{s}$ & Sing.\\
   \hline
   $(-\infty,-2)$ &  finite &   0 & 0 & $\infty$ & IV or V\\ 
   $-2$ & finite &    0 & finite & $\infty$ & IV \\
   $(-2,-1]$ & finite &  0 &  $\infty$ & $\infty$ & II  \\
   $(-1,0)$, $K_{1}\neq0$  & finite 
      & finite & $\infty$ & $\infty$ &  II 
      \\
   $(-1,0)$, $K_{1}=0$ & finite 
      & $\infty$ & $\infty$ & $\infty$ & III \\
0 &  0/$\infty$ & $\infty$ & $\infty$ & finite & big crunch / rip \\
   $(0,\infty)$ &0/$\infty$&   $\infty$ & $\infty$ & -1 & grand 
   crunch / rip \\
   \hline
   \end{tabular}
\caption{Singularities in terms of the power expansion of $q(t)$}
\label{tablw}
\end{table}

After considering the case of finite $t_{0}$ we are not to forget the 
infinite case. For this we need asymptotic expressions for our 
magnitudes,
\begin{equation}
a(t)=\exp\left(\int_{t}^\infty \left(\int_{t''}^\infty 
h(t')\,dt'+K_{1}\right)^{-1}dt''\right),
\end{equation}
\begin{equation}
\rho(t)=3\left(\int_{t}^\infty h(t')\,dt' +K_{1}\right)^{-2},
\end{equation}
\begin{equation}
p(t)=\frac{2h(t)-3}{\left(
\displaystyle\int_{t}^\infty h(t')\,dt'+K_{1} \right)^{2}}\end{equation}
which require $K_{1}=0$ for $\rho$ and $p$ to diverge at infinity. 

Again, calculations are involved and the results are summarised in
Table~\ref{tablinf}, where the asymptotic behavior of $h(t)$ for large
$t$ is related to the asymptotic values of the scale factor, the
energy density, the pressure and the barotropic index and to the type
of singularity or future behavior according to our classifications. 
Again, the last column refers to the thorough classification of the 
singularities.
\begin{table}[h]
   \begin{tabular}{ccccccccc}
   \hline
   $h$ & signum\,($h$) & $K_{1}$ &$a_{\infty}$ & $\rho_{\infty}$ &$p_{\infty}$ & $w_{\infty}$ & 
   Behavior\\
   \hline Finite
$\int_{t}^\infty h(t')\,dt'$ & + & 0& 0 &   $\infty$ &  $\infty$ & -1 &  $\infty$
\\ & - & 0& $\infty$ &   $\infty$ &  $\infty$ & -1 &  little rip / sibling
\\  & $\pm$ & positive & 0 &   finite &  finite & -1 &  non-singular
\\  & $\pm$ & negative & $\infty$ &   finite &  finite & -1 &  pseudo-rip
\\ $t^{-1}\lesssim |h(t)|\to 0$ & +& any  & $\infty$ &   0 &  0 & -1 & little 
rip with 0 $\rho$ and $p$ 
\\  & - & any& 0 &   0 &  0 & -1 &  $\infty$ 
\\ $K$ & +& any & $\infty$ &   0 &  0 & -1+2K/3 &  non-singular
\\ $K\in (-1,0)$ &- & any &  0 &   0 &  0 & -1+2K/3 &  $\infty$
\\$K\in  (-\infty,-1]$ & -& any  & 0 &   0 &  0 & -1+2K/3 & non-singular 
\\   $|h(t)|\to \infty$, \ infinite $\int_{t}^\infty dt''/\int_{t''}^\infty h(t')\,dt'$& + & any& 
$\infty$ &   0 &  0 & $\infty$ & non-singular 
\\   $|h(t)|\to \infty$, \ infinite $\int_{t}^\infty dt''/\int_{t''}^\infty h(t')\,dt'$& - 
& any& 0 &   0 &  0 & $\infty$ & non-singular 
\\   $|h(t)|\to \infty$, \ finite $\int_{t}^\infty dt''/\int_{t''}^\infty h(t')\,dt'$& $\pm$ & any& finite &   0 &  0 & $\infty$ & non-singular 
\\  
\hline
   \end{tabular}
\caption{Possible behavior in terms of the asymptotic behavior of 
$q(t)$}\label{tablinf}
\end{table}

\section{Singularities in cosmological models}

So far we have been able to detect all possible singular scenarios in 
spatially flat cosmological models by use of power expansions in the 
time coordinate. The resulting possible singularities which have been 
obtained may be organised in the following fashion with the names and 
references where they were discovered for the first time:

\begin{itemize}    
\item Type -1: ``Grand bang/rip'': \cite{grandrip} The scale factor
vanishes or blows up at $w=-1$.  The Hubble ratio, the energy density
and the pressure blow up.  These are strong singularities. 

We may see as the counterparts for Big Bang and Big Rip singularities
$w\neq-1$ at the phantom divide, but pressure and energy density blow
up as a power of $t$ different from -2.  

The sign of the coefficient $h_{0}$ states the kind of singularity: if
$h_{0}>0$, we have a sort of exponential Big Bang singularity (Grand
bang or Grand crunch if we exchange $t$ for $-t$) and the phantom
divide is approached from below.  If $h_{0}<0$, the scale factor blows
up and we would have an exponential Big Rip at $t=0$ (Grand Rip).  The
phantom divide is approached from above.
   
\item Type 0: ``Big bang/crunch'': These are well known
\emph{classical} strong singularities. The Hubble ratio, the energy density
and the pressure diverge.
   
\item Type I: ``Big rip'' \cite{Caldwell:2003vq}: These were the
\textit{new} singularities to appear in cosmological models.  The
scale factor blows up instead of vanishing. In this sense, the final 
fate of the universe would be a progressive disruption of all 
structures instead of a collapse. These are strong singularities. A 
peculiar feature is that lightlike geodesics are complete close to 
the singularity.

\item Type II: ``Sudden singularities'' 
\cite{sudden0,sudden,sudden1,sudden2,sudden3,sudden4,sudden5,sudden6,sudden7,sudden8,
sudden9,sudden10,sudden11,sudden12} or even
``quiescent singularities''\cite{quiescent}: This is the second type
of singularity that was considered on introducing new cosmological
models, though they had been already introduced in \cite{suddenfirst}.
The main feature of these singularities is that the scale factor, the
Hubble ratio and the energy density do not blow up and the models just
violate the dominant energy condition.  The appearance of fractional
exponents in the power expansion of the scale factor produces
divergent derivatives starting from second onwards.  These
singularities are weak and in this sense they cannot be considered the end of the universe
\cite{suddenferlaz}, since the spacetime can be extended beyond the 
singularity.  In certain contexts they have been named
\cite{brake} or big boost \cite{boost}.

\item Type III: ``Big freeze'' \cite{freeze} or ``finite scale factor
singularities'': Similar to the previous one, but the first divergent
derivative of the scale factor is the first and hence the Hubble
factor, the energy density and the pressure blow up.  That is, the
first derivative of the scale factor is singular.  Depending on the
definition used \cite{tipler,krolak}, they can be either strong or
weak \cite{puiseux}.

\item Type IV \cite{tsagas}: A generalisation of sudden singularities
for models with a divergent derivative of the scale factor of order
higher than two and therefore, pressure and density do not blow up.
They are also weak singularities.
   
\item Type V: ``$w$-singularities'' \cite{wsing, loitering}: They 
could be considered a subcase of the previous ones, since every 
derivative of the scale factor is finite and just the barotropic 
index $w$ blows up.  They are weak singularities \cite{barotrope}.
   
\item Type $\infty$: ``Directional singularities'' \cite{hidden,initial}:
These may come up at infinite coordinate time, though finite proper
time.  They do not affect all geodesics, since comoving observers need
infinite proper time to reach the singularity.  In this sense they are
directional and are p.p. curvature singularities, but they are strong.
\end{itemize}

In addition to these singularities, there are other asymptotic
behaviors which resemble singular ones, though they correspond to
regular models.  For all of them the barotropic index is near the
phantom divide and could be deemed deviations from $\Lambda$-Cold
Dark Matter model:

\begin{itemize}
\item Little Rip  \cite{little,little1}: The Hubble ratio diverges at infinite
coordinate and proper time. 

\item  Little Sibling of the Big Rip \cite{sibling}: A subcase of the 
previous behavior, but with a finite derivative of the Hubble 
ratio.

\item Pseudo-rip: \cite{pseudo} Asymptotic monotonic, finite growth of
the Hubble ratio.

\end{itemize}

\section{Conclusion}

We have provided a thorough classification of possible final fates 
for our universe in terms of power and asymptotic
expansions of the deceleration parameter of the models. This provides 
a unified framework for both singular and non-singular behaviors. 
Besides, we have analysed if the singularities are strong or weak. In 
the latter case, they cannot be considered the final stage for the 
universe and the spacetime can be continued beyond the singularity.

Among the main interesting features we may point out that the popular
sudden and generalised sudden singularities are not a final fate in
the sense they are weak singularities and the spacetime can be 
extended beyond the singularity.  And that intriguing
directional singularities may come up.  That is, singularities which
may be experienced by some observers, but not by all of them, though
they are strong.  Besides, Big Rip singularities are avoided by
photons.

\vskip6pt

\enlargethispage{20pt}

\ethics{Insert ethics statement here if applicable.}

\dataccess{Insert details of how to access any supporting data here.}

\aucontribute{For manuscripts with two or more authors, insert details of the authors’ contributions here. This should take the form: 'AB caried out the experiments. CD performed the data analysis. EF conceived of and designed the study, and drafted the manuscript All auhtors read and approved the manuscript'.}

\competing{Insert any competing interests here. If you have no competing interests please state 'The author(s) declare that they have no competing interests’.}

\funding{Insert funding text here.}

\ack{Insert acknowledgment text here.}

\disclaimer{Insert disclaimer text here if applicable.}



\begin{thebibliography}{9}
	
\bibitem{geroch} Geroch R. 1968.
What is a singularity in general relativity?
\textit{Annals of Physics} \textbf{48}, 526.

\bibitem{HE} Hawking SW, Ellis GFR. 1973.  
\textit{The Large Scale
Structure of Space-time}, Cambridge, UK: University Press.

\bibitem{beem} Beem J, Ehrlich P. 1981.
\textit{Global Lorentzian Geometry}, New York, USA: Dekker.

\bibitem{visser} Catto\"en C, Visser M. 2005.  Necessary and
sufficient conditions for big bangs, bounces, crunches, rips,
sudden singularities, and extremality events.
\textit{Class.\ Quant.\ Grav.}\  {\bf 22}, 4913.

\bibitem{puiseux} Fern\'andez-Jambrina L, Lazkoz R. 2006.
Classification of cosmological milestones. 
\textit{Phys.\ Rev.\  D} {\bf 74}, 064030.

\bibitem{modigravi} Fern\'andez-Jambrina L, Lazkoz R. 2009.
Singular fate of the universe in modified theories of gravity. 
\textit{Phys. Lett. B} \textbf{670},  254.

\bibitem{hidden} Fern\'andez-Jambrina L. 2007. 
Hidden past of dark energy cosmological models.
\textit{Phys.\ Lett.\ B} \textbf{656}, 9.

\bibitem{grandrip} Fern\'andez-Jambrina L. 2014.
Grand rip and grand bang/crunch cosmological singularities.
\textit{Phys.  Rev.  D} \textbf{90},  064014.

\bibitem{initial} Fern\'andez-Jambrina L. 2016.
Initial directional singularity in inflationary models. 
\textit{Phys.  Rev.  D} \textbf{94}, 024049.

\bibitem{ellis} Ellis GFR., Schmidt BG. 1977.
Singular space-times. 
\textit{Gen. Rel. Grav.} \textbf{8}, 915 (1977).

\bibitem{tipler} Tipler FJ. 1977 
Singularities in conformally flat spacetimes.
\textit{Phys. Lett. A} \textbf{64}, 8. 

\bibitem{krolak} Kr\'olak A. 1986.
Towards the proof of the cosmic censorship hypothesis. 
\textit{Class. Quant. Grav.} \textbf{3}, 267. 


\bibitem{clarke} Clarke CJS, Kr\'olak A. 1985.
Conditions for the occurence of strong curvature singularities.
\textit{Journ. Geom. Phys.} \textbf{2}, 127. 

\bibitem{Nojiri:2005sx}
Nojiri S, Odintsov SD, Tsujikawa S. 2005
Properties of singularities in (phantom) dark energy universe.
\textit{Phys.\ Rev.\  D} {\bf 71}, 063004.

\bibitem{IV} D\c abrowski MP, Marosek K. 2013.
Standard and exotic singularities regularized by varying constants.
\textit{J. Cosmol. Astropart. Phys.} \textbf{13}, 012.

\bibitem{yurov} Yurov AV. 2010.
Brane-like singularities with no brane.
\textit{Phys. Lett. B} \textbf{689}, 1.

\bibitem{sesto} D\c abrowski MP, Marosek K, Balcerzak, A. 2014.
\textit{Memorie della Societa Astronomica Italiana} \textbf{85}, 44.

 


\bibitem{Caldwell:2003vq}
Caldwell RR, Kamionkowski M, Weinberg NN. 2003.
Phantom Energy and Cosmic Doomsday.
\textit{Phys.\ Rev.\ Lett.}\  {\bf 91} 071301.

\bibitem{sudden0} Barrow JD. 2004.
Sudden future singularities. 
\textit{Class. Quant. Grav.} {\bf 21}, L79

\bibitem{sudden} Nojiri S, Odintsov SD.  2004
Quantum escape of sudden future singularity. 
\textit{Phys.\ Lett. B}  {\bf 595}, 1.

\bibitem{sudden1}
Barrow JD 2004.
More general sudden singularities. 
\textit{Class.\ Quant.\ Grav.}\  {\bf 21}, 5619.

\bibitem{sudden2}
Lake K. 2004.
Sudden future singularities in FLRW cosmologies. 
\textit{Class.\ Quant.\ Grav.}  {\bf 21}, L129.

\bibitem{sudden3}
Nojiri S, Odintsov SD. 2004.
Final state and thermodynamics of a dark energy universe. 
\textit{Phys.\   Rev.\ D} {\bf 70}, 103522

\bibitem{sudden4}  D\c abrowski MP. 2005
Inhomogenized sudden future singularities.  
\textit{Phys.\ Rev.\ D} {\bf 71}, 103505.

\bibitem{sudden5}
Chimento LP,  Lazkoz R. 2004.
On big rip singularities. 
\textit{Mod.\ Phys.\ Lett.\ A} {\bf 19}, 2479.

\bibitem{sudden6}
D\c abrowski MP. 2005.
Statefinders, higher-order energy conditions, and sudden future 
singularities. 
\textit{Phys.\ Lett.\ B} {\bf 625}, 184.

\bibitem{sudden7}
Barrow JD, Batista AB, Fabris JC, Houndjo S. 2008 
Quantum particle production at sudden singularities. 
\emph{Phys. Rev. D} \textbf{78}, 123508.

\bibitem{sudden8} Barrow JD, Lip SZW. 2009.
Classical stability of sudden and big rip singularities. 
\emph{Phys. Rev. D} \textbf{80}, 043518.

\bibitem{sudden9} Nojiri S, Odintsov SD. 2008.
Future evolution and finite-time singularities in F(R) gravity 
unifying inflation and cosmic acceleration. 
\emph{Phys. Rev. D} \textbf{78}, 046006.

\bibitem{sudden10} Barrow JD, Cotsakis S,  Tsokaros. 2010.
The Construction of Sudden Cosmological Singularities. 
\emph{Class. Quant. Grav.} \textbf{27}, 165017.

\bibitem{sudden11}
Singh P. 2012.
Curvature invariants, geodesics, and the strength of singularities in 
Bianchi-I loop quantum cosmology. 
\emph{Phys. Rev. D} \textbf{85}, 104011.

\bibitem{sudden12}
Denkiewicz T,  D\c abrowski MP,  Ghodsi H, Hendry MA. 2012.
Cosmological tests of sudden future singularities. 
\emph{Phys. Rev. D} \textbf{85}, 083527.

\bibitem{quiescent} Shtanov Y, Sahni V. 2002. 
New cosmological singularities in braneworld models. 
\textit{Class.  Quant.  Grav.} \textbf{19}, L101.

\bibitem{suddenfirst}  Barrow JD, Galloway GJ, Tipler FJ. 1986. 
The closed-universe recollapse conjecture. 
\textit{MNRAS} \textbf{223}, 835.

\bibitem{suddenferlaz} Fern\'andez-Jambrina L, Lazkoz R. 2004.
Geodesic behaviour of sudden future singularities. 
\textit{Phys.\ Rev.\  D} {\bf 70}, 121503.

\bibitem{brake} Gorini V, Kamenshchik AY,  Moschella U, 
Pasquier V. 2004. 
Tachyons, scalar fields, and cosmology. 
\emph{Physical Review D} {\bf 69}, 123512.
 
\bibitem{boost} Barvinsky AO, Deffayet C, Kamenshchik AY. 2010.
CFT driven cosmology and the DGP/CFT correspondence. 
\textit{JCAP} \textbf{05}, 034.

\bibitem{freeze} Bouhmadi-L\'opez M, Gonzalez-D\'\i az PF, and
Mart\'\i n-Moruno P. 2008. 
Worse than a big rip? 
\textit{Phys.\ Lett.\  B} {\bf 659}, 1.


 
\bibitem{tsagas} Barrow JD, Tsagas CG. 2005. 
New Isotropic and Anisotropic Sudden Singularities 
\textit{Class.\ Quant.\ Grav.}\  {\bf 22}, 1563.

\bibitem{wsing} D\c abrowski MP, Denkiewicz T. 2009. 
Barotropic index w-singularities in cosmology. 
\textit{Phys. Rev. D}  \textbf{79}, 063521.

\bibitem{loitering} Shtanov Y, Sahni V. 2005. 
Did the universe loiter at high redshifts? 
\emph{Phys.  Rev.  D} \textbf{71}, 084018.
 
\bibitem{barotrope} Fern\'andez-Jambrina L. 2010. 
w-cosmological singularities.
\textit{Phys. Rev. D} \textbf{82}, 124004.

\bibitem{little} Frampton PH, Ludwick KJ, Scherrer RJ. 2011. 
The little rip.
\textit{Phys. Rev.  D} \textbf{84}, 063003;

\bibitem{little1} Frampton PH,  Ludwick KJ,  Nojiri S, Odintsov SD, Scherrer RJ. 2012.
Models for little rip dark energy. 
\textit{Phys.  Lett.  B} \textbf{708}, 204.

\bibitem{sibling} Bouhmadi-Lopez M, Errahmani A, Martin-Moruno P, 
Ouali T, Tavakoli Y. 2014.
The little sibling of the big rip singularity.
\textit{International Journal of Modern Physics D} \textbf{24}, 1550078.

\bibitem{pseudo} Frampton PH,  Ludwick KJ, Scherrer RJ 2012.
Pseudo-rip: Cosmological models intermediate between the cosmological 
constant and the little rip.
\textit{Phys.  Rev.  D} \textbf{85}, 083001.

%
%
%
%

\end{thebibliography}
\end{document}